\documentclass[12pt,preprint]{aastex}

\newcommand{\kms}{\ensuremath{\,\hbox{km s}^{-1}}}
\newcommand{\msun}{\ensuremath{M_\odot}}
\newcommand{\ml}{\ensuremath{M/L}}
\newcommand{\cv}{\ensuremath{\epsilon}}

\begin{document}

\title{On the Correlations of Massive Black Holes with their Host Galaxies}

\author{Gregory S. Novak,\altaffilmark{1} S.
  M. Faber,\altaffilmark{1} and Avishai Dekel\altaffilmark{2}}

\altaffiltext{1}{UCO/Lick Observatories, University of California,
Santa Cruz, CA 95064}
\altaffiltext{2}{The Hebrew University, Jerusalem 91904, Israel}

\shorttitle{BLACK HOLE MASS CORRELATIONS}
\shortauthors{NOVAK, FABER, and DEKEL}

\begin{abstract}
  We address the correlations of black hole (BH) mass with four
  different host-galaxy properties from 11 existing data sets.
  For the purpose of guiding theoretical understanding, we first try
  to quantify the tightness of the intrinsic correlations.  We assume
  that all of the relations are power laws and perform linear
  regressions that are symmetric in the two variables on the logarithms
  of the data points.  Given the estimated measurement errors, we
  evaluate the probability distribution of the residual variance in
  excess of that expected from the measurement errors.  Our central
  result is that the current data sets do not allow definite
  conclusions regarding the quality of the true correlations because
  the obtained probability distributions for the residual variance
  overlap for most quantities.  Velocity dispersion as collected by
  Merritt and Ferrarese ($\sigma_{\rm MF}$) and galaxy light concentration as
  measured by Graham and coworkers ($C_{\rm Re}$) are consistent with zero
  residual variance.  Taken at face value, this means that these two
  correlations are better than the others, but this conclusion is
  highly sensitive to the assumed measurement errors and would be
  undone if the present estimated errors were too large.  We then
  consider which of the relations offer the best inferences of BH mass
  when there is no direct measurement available.  As with the residual
  variances, we find that the probability distribution of expected
  uncertainty in inferred BH masses overlaps significantly for most of
  the relations.  Photometric methods would then be preferred because
  the data are easier to obtain, as long as bulge-disk decomposition
  or detailed modeling of the photometric profile (as studied by 
  Graham and coworkers do not present problems.  Determining which
  correlation offers the best inferences requires reducing the
  uncertainty in the expected error in the inferred BH masses (the
  ``error on the error'').  This uncertainty is currently limited by
  uncertainty in the residual variance for all of the relations.  The
  only quantities for which BH mass inferences are limited by
  measurement error are $\sigma_{\rm MF}$ and $C_{\rm Re}.$ Therefore, if
  these relations are {\em truly} better than the others, then new,
  improved measurements should allow improved inferences of BH masses.
  If they do not, the conclusion must be that the present low residual
  variances for these two relations result from overestimated error
  bars.
  \end{abstract}


\maketitle

\section{Introduction}
There is evidence for the presence of massive BHs in the centers of
early-type galaxies \citep[e.g.,][]{kormendy:95}.  Spectroscopic and
photometric data of high spatial resolution from the {\em Hubble Space
Telescope} have made it possible to derive relatively accurate BH
masses for a number of nearby galaxies \citep{kormendy:01}. This has
led to empirical derivations of relationships between the BH mass and
several different properties of its host galaxy, such as bulge mass
\citep{magorrian:98, haering:04}, stellar velocity dispersion
\citep{gebhardt:00, ferrarese:00}, bulge luminosity
\citep{kormendy:93, mclure:02,marconi:03, bettoni:03}, and galaxy light
concentration \citep{graham:01}.

These correlations pose a theoretical challenge because
the mass accretion onto BHs takes place on extremely small spatial
scales compared to the scales corresponding to these global properties
of galaxies.  For the purpose of constructing a theoretical model
explaining these correlations, it is essential to establish which of
these correlations has the smallest intrinsic residual variance, and
therefore has the best chance of being causally linked to the BH mass.
This is a difficult problem because the global properties are
correlated with each other, as well as with the central BH mass.

We analyze the correlations between BH mass and several global
galactic properties as described in detail in \S\ref{sec:data}.
An interesting quantity for comparing tightness of the correlations is
the ``residual'' variance, which is the excess variance in the
correlation that is not explained by the observational errors.  The
residual variance reflects either the true scatter due to hidden
variables other than the two variables whose correlation is analyzed
(what one might call ``intrinsic'' variance) {\em or} spurious
scatter in the correlation due to underestimates of
observational or systematic errors.  We posit that the most physically
meaningful correlation is that with the smallest residual variance.
Identifying that correlation is termed the ``Theorist's Question.''
In addition to inferring the most likely value of the residual
variance given the data, we also consider the {\em uncertainty} in our
inferences about this quantity.

A second question is which correlation allows the most accurate 
{\em inferences} of BH mass when a direct measurement of it is
unavailable.  We term this the ``Observer's Question,'' which is
distinct from the Theorist's because the quantity serving as a
predictor of BH mass has observational errors.  For example, the
correlation with a given quantity could be the theorist's favorite,
with zero residual variance, while being useless for inferring BH
masses because of the large errors involved in measuring it.
Observers would also like the quantity under consideration to be easy
to measure in terms of telescope time.  Such considerations do not
enter into the statistical part of the problem, but it may be
desirable lose a small amount of accuracy to save a large amount of
observational effort.

Section \ref{sec:method} describes our method of analysis, 
\S\ref{sec:data} discusses the data sets upon which this analysis is
based, and \S\ref{sec:results} contains a discussion of our
results.  Finally, \S\ref{sec:conclusions} presents a review of
our conclusions.

\section{Method}
\label{sec:method}
The data in each case consist of $N$ points $\{x_i,y_i\}$, where $y_i$
is the base-10 logarithm of the measured BH mass and $x_i$ is the
base-10 logarithm of the measured galaxy quantity (in the case of
velocity dispersion or mass) or the quantity itself (in the case of
magnitude or light concentration).  Also available are the associated
1 $\sigma$ measurement error bars on each quantity, $\sigma_{yi}$ and
$\sigma_{xi}$.  We assume Gaussian errors.

We assume that all of the relations between the logarithms are linear
(power laws).  If one relation has a functional form more complex than
a power law, our model will assign it an unfairly high intrinsic
variance.  A model more complex than a power law may decrease the
intrinsic variance of a given correlation, but it will also involve
a larger number of model parameters.

The objects in each data set are nearby galaxies with central BH
masses derived from stellar kinematics, gas dynamics, or masers.  Some
data sets include galaxies of all Hubble types, while others are
restricted to early types.  See \S\ref{sec:data} for details on
the individual samples.

\subsection{Theorist's Question}
\label{sec:method-theorist}
We first seek to compare estimates of the residual variance ($\cv^2$)
resulting from the correlation of several different galactic quantities with
BH mass.  We wish to find the quantity that yields the smallest
residual variance, and therefore the tightest correlation.  The hope
is that the residual variance is the ``intrinsic'' variance.

As pointed out in \citet{tremaine:02} and elsewhere, in the absence of
information about the direction of the causal link between BH mass and
other galactic properties, there is no natural division of the
variables into ``dependent'' and ``independent'' variables. Therefore,
$x$ and $y$ are treated as symmetrically as possible.

In extensive Monte Carlo simulations (G. S. Novak et al. 2006, in
preparation) to investigate the performance of many fitting methods with
small numbers of data points, errors in both coordinates, and the
presence of residual variance, we found that the algorithm defined by
\citet{tremaine:02} based on FITEXY \citep{press:92} estimates the
slope with the least bias and variance.   Our results agree with those
of \citet{tremaine:02}, that the competing estimator defined by
\citet{akritas:96} gives reliable estimates of the slope {\em only} if
the spread of $x$-values is large compared to their errors and if the
$x$- and $y$-errors of all points are comparable.  If the spread of the
$x$-values is too small compared to the $x$-errors, then the
\citet{akritas:96} estimator develops a bias toward large slopes.  If
some points have error bars that are much larger than the other
points, then the \citet{akritas:96} estimator becomes inefficient.  A
statistical estimator is inefficient if there is another estimator
with smaller variance \citep{cowan:98}.

The same Monte Carlo simulations also confirmed that, even using the
extension of the FITEXY estimator defined by \citet{tremaine:02}, it
is unfortunately impossible to avoid specifying about whether the
residual variance is in the $x$-coordinate or in the $y$-coordinate.
Assuming that the residual variance is in the $y$-coordinate when in
truth it is in the $x$-coordinate results in systematically low
slopes, and vice versa.  

The Monte Carlo study included unweighted least squares fitting, weighted
least squares fitting, weighted orthogonal regression (as defined by the
FITEXY routine), the modified FITEXY routine defined by
\citet{tremaine:02}, the estimator defined by \citet{akritas:96}, a
Bayesian estimator defined by \citet{gull:89}, and a Bayesian
estimator we defined using an unfinished manuscript by 
\citet[unpublished]{jaynes:91}.\footnote{See
  http://bayes.wustl.edu/etj/articles/leapz.pdf.}
If a given estimator treats $x$ and $y$ 
asymmetrically, we considered the same estimator with $x$ and $y$
interchanged.  Also included were the lines defined by the arithmetic,
geometric, and harmonic means of these two slopes, as well as the
bisector line.  The modified FITEXY estimator gave the most efficient
and unbiased estimates of the slope when the fitting method
incorporated the residual variance into the correct coordinate.  When
the fitting method includes residual variance in the wrong coordinate,
the slope found by all estimators is biased as noted above
(G. S. Novak et al. 2006, in preparation).

Therefore, we use the FITEXY estimator defined by \citet{press:92} as
modified by \citet{tremaine:02} to include residual variance.  That
is, for a given value of \cv\, we find the values of $\alpha$ and
$\beta$ that minimize
\begin{equation}
  \chi(\alpha, \beta)^2 = 
  \sum_{i=0}^N 
    \frac{[y_i-\beta (x-x_0) - \alpha]^2}
         {\sigma_{yi}^2 + \beta^2 \sigma_{xi}^2 + \cv^2 [c + (1-c) \beta^2]},
\label{chi2}
\end{equation}
where $\alpha$ and $\beta$ are the offset and slope of the
correlation, respectively, $x_0$ is a value chosen near the mean of
the distribution of $x$-values in order to reduce covariance between
$\alpha$ and $\beta$, and $c=1$ if the residual variance is assumed to be
in the $y$-coordinate while $c=0$ if it is assumed to be in the
$x$-coordinate.  Note that \cv\ will have different units depending on
the value of $c$.
If the reduced $\chi^2$ of the fit is not equal to 1, we
adjust \cv\ and refit until it is.  Finally, to estimate the upper and
lower limits on \cv, we adjust \cv\ and refit until the reduced
$\chi^2$ is equal to $1 \pm (2/N)^{1/2}$.  This gives an estimate of
the $1 \sigma$ error bar on \cv.

The fact that one must specify the coordinate containing the residual
variance complicates comparisons to theoretical
models.  If one develops a theoretical model in which the galaxy velocity
dispersion regulates BH growth, such as \citet{adams:01}, then the
residual variance would effectively lie in the BH mass.  On the other
hand, in a theoretical model in which the BH profoundly affects the
structure of its host galaxy through quasar activity, such as
\citet{silk:98}, the residual variance would lie in the
galaxy property.  Therefore, theoretical models should attempt to match
{\em different} slopes depending on the structure of the model.

However, blind comparisons to the different slopes can run into
problems because of observational selection effects. Galaxies are
selected for BH mass studies based on their properties, not their
as-yet-unknown BH mass, so regarding the BH mass as the independent
variable for a linear fit will result in selection criteria that are
difficult to quantify.  For example, if galaxies are selected based on
a hard limit in the galaxy property and one then regards the BH mass
as the independent variable, the slope will be systematically low.

The best procedure would be for theoretical models to provide the full 
probability distribution linking BH mass and other galactic
properties.  It would then be possible to make any desired comparison
to observational data while properly accounting for selection effects.
Unfortunately, theoretical models for BH formation are generally
far from this level of precision.

The answer to the Theorist's Question is extremely sensitive to
inaccurate estimates of measurement errors.  If measurement errors are
underestimated, then the fit will find too much residual variance.
Similarly, if the measurement errors are overestimated, the fit will
find too little residual variance.  Finally, if an observer
overestimates measurement errors so severely that the correlation
becomes ``too good,'' in the sense that the observed scatter is 
{\em less} than that expected from the error bars, the result will be a
\cv\ distribution sharply peaked around zero, and the reduced $\chi^2$
of the linear fit will be less than 1 even when \cv\ is set to zero.

\subsection{Observer's Question}
\label{sec:method-observer}
Next, we wish to quantify the predictive power of each correlation.
If we had a probability distribution for the model
parameters, the machinery of Bayesian statistics could be used to
construct the predictive distribution for BH masses given a predictor
variable.  This is the probability distribution for an
about-to-be-measured $y$-value given a just-measured $x$-value and
model parameters that are constrained by all of the 
previously collected data.  The important thing about the predictive
distribution in Bayesian statistics is that it includes {\em both}
uncertainty in the model parameters {\em and} expected observational
uncertainty\footnote{It is assumed that the errors in the
  just-measured $x$-value are statistically typical of previously
  measured $x$-values in the data set.  In other words, the quality of
all $x$-measurements is the same.}.

In the Bayesian formalism, usually one makes a $\chi^2$ estimator into
a probability distribution for the model parameters via $p(M|D)
\propto p(M) \exp(-\chi^2(M,D)/2)$, where $\chi^2$ is given by equation
\ref{chi2}, $M$ is a set of model parameters, and $D$ is a set of
data.  Unfortunately, allowing observational errors in both $x$ and
$y$ makes the distribution for the model parameters depend
nontrivially on the prior distribution of the unobservable parameter
(BH mass in our case) in a manner that does not become less important
as the number of data points grows (as is usually the case; 
\citet{gull:89}).  Therefore, the standard Bayesian predictive
distribution is not available, but we can construct a ``poor man's''
estimate of the desired quantity---the uncertainty in inferences about
$y$ given both observational uncertainty in $x$ and uncertainty in the
model parameters---by combining variances in the standard way.  The
expected variance is
\begin{equation}
  \sigma_y^2 = (x-x_0)^2 \sigma_\beta^2 + \beta^2 \sigma_x^2 +
  \sigma_\alpha^2 + [c + (1-c) \beta^2] \cv^2,
  \label{eq:predictive}
\end{equation}
where $\sigma_y$ is the expected uncertainty in a ``new'' $y$-value,
$\sigma_x$ is the expected observational uncertainty in a new
$x$-value, and $\sigma_\beta$ and $\sigma_\alpha$ are the
uncertainties in the slope and offset given the data collected thus far.

It is important to realize that the answer to the Observer's Question
is far less sensitive to inaccurate estimates of measurement errors
than the answer to the Theorist's Question.  Either over- or
underestimating the error bars will mislead the theorist, whose main
interest is residual variance, but the {\em total} uncertainty
resulting from the combination of residual variance and measurement
error will remain virtually unchanged.  Therefore, the quality of the
observer's inferences about BH masses will be unchanged.

Distance uncertainty has a different impact on the Theorist's and
Observer's questions.  The theorist would prefer a correlation where
both variables scale the same way with distance, such as BH mass and
bulge mass.  This will preserve a tight correlation even when the
distances to the individual objects are uncertain.  The observer would
prefer a quantity that does not scale with distance, such as velocity
dispersion on galaxy light concentration, so that BH mass inferences
will not depend sensitively on the unknown distance to the object.
Thus, the theorist will have an easy time calibrating a tight relation,
but a hard time using that relation to infer BH masses.  The observer
will have a hard time calibrating a relation (because distances must
be known accurately), but an easier time inferring BH masses after the
calibration is finished.

\section{Data Sets}
\label{sec:data}
We analyze several observable galaxy properties that correlate with
BH mass.  Although reverberation mapping studies in active galactic
nuclei are rapidly converging to the existing BH correlations for
inactive local galaxies in terms of residual variance
\citep{onken:04}, in order to maintain homogeneity they are excluded
from this study.  We use the BH masses and distances quoted in each
paper with the exception of NGC 821, for which we revise the BH mass
estimate as indicated by \citet{richstone:04}, scaling to a different
distance if necessary.

Our core sample is that of \citet{gebhardt:03} because this sample
consists of 10 E and 2 S0 galaxies with BH masses and \ml\ values all
inferred from similar dynamical models of the galactic potential.  The
sample is therefore homogeneous.  The derived bulge masses do not
suffer from large perturbations due to the presence of small disks.

The velocity dispersions are denoted $\sigma_{\rm G}$.  Following
\citet{tremaine:02} we assume 5\% errors on the velocity dispersions.

\begin{figure}[htbp]
  \includegraphics[height=.25\textwidth]{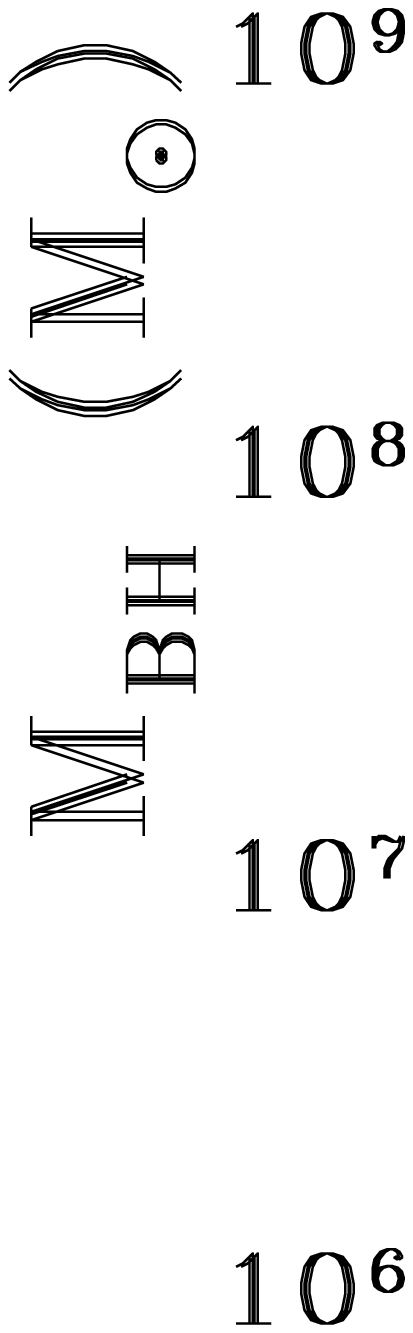}  
  \includegraphics[height=.25\textwidth]{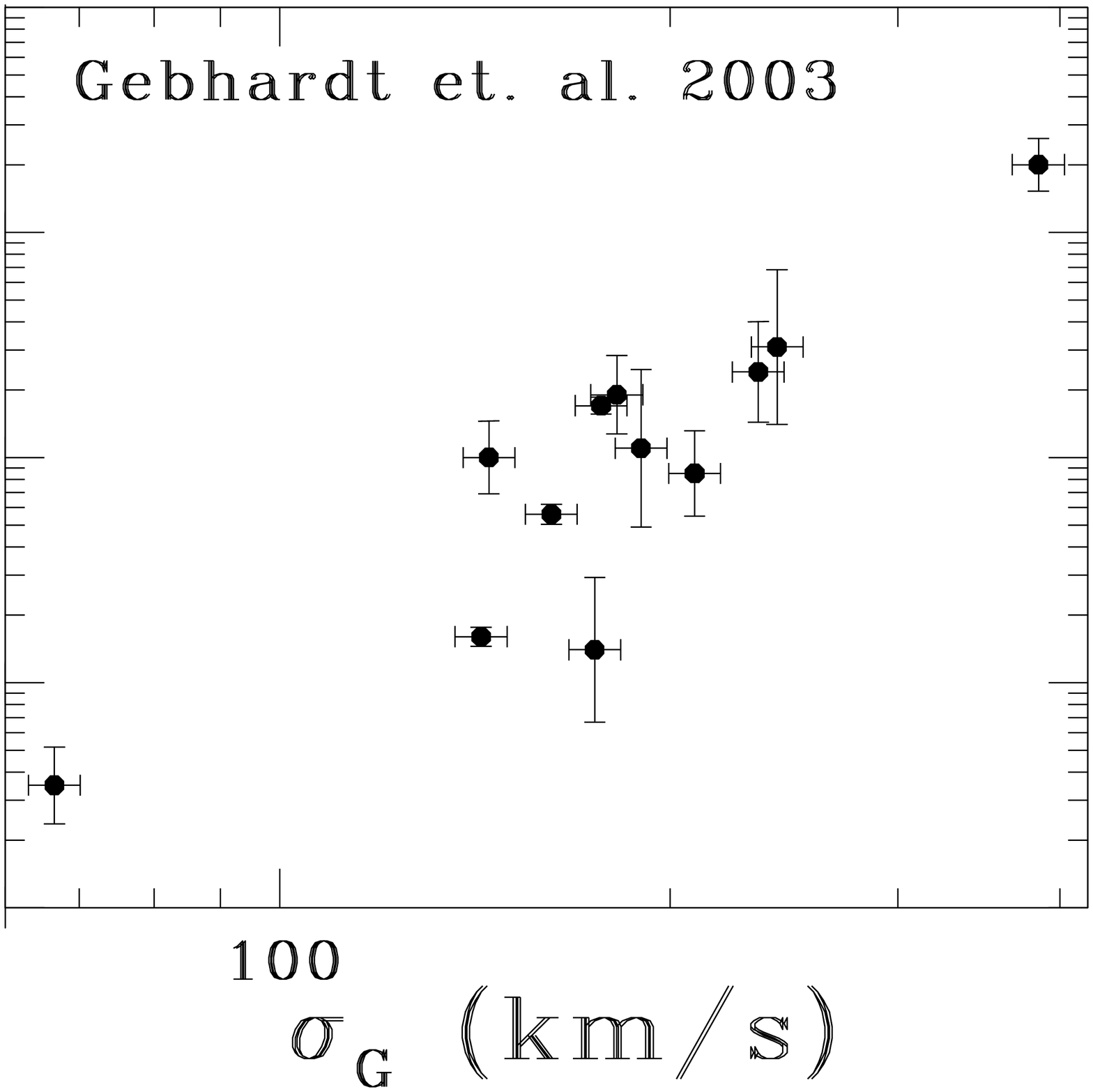}  
  \includegraphics[height=.25\textwidth]{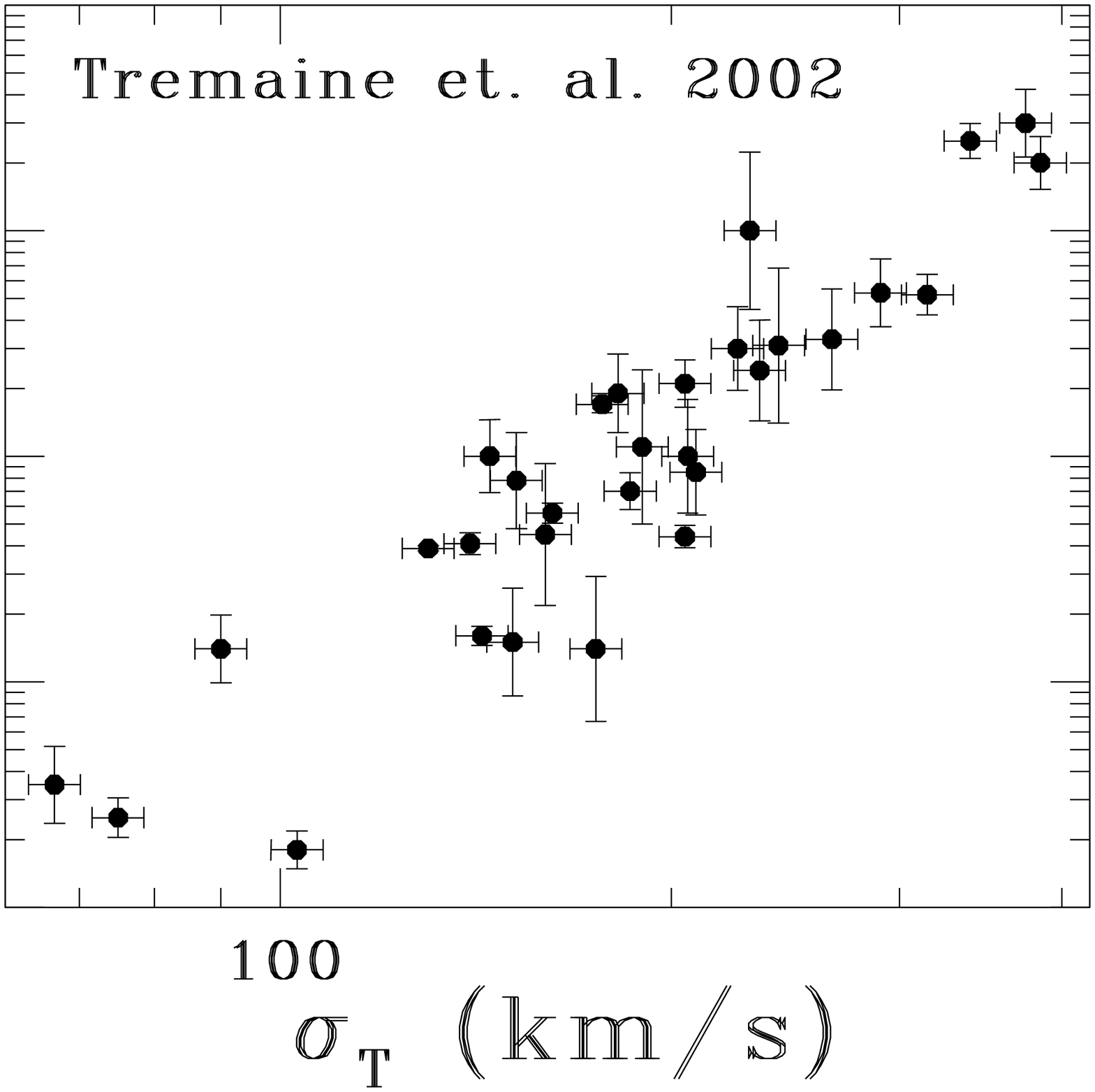}  
  \includegraphics[height=.25\textwidth]{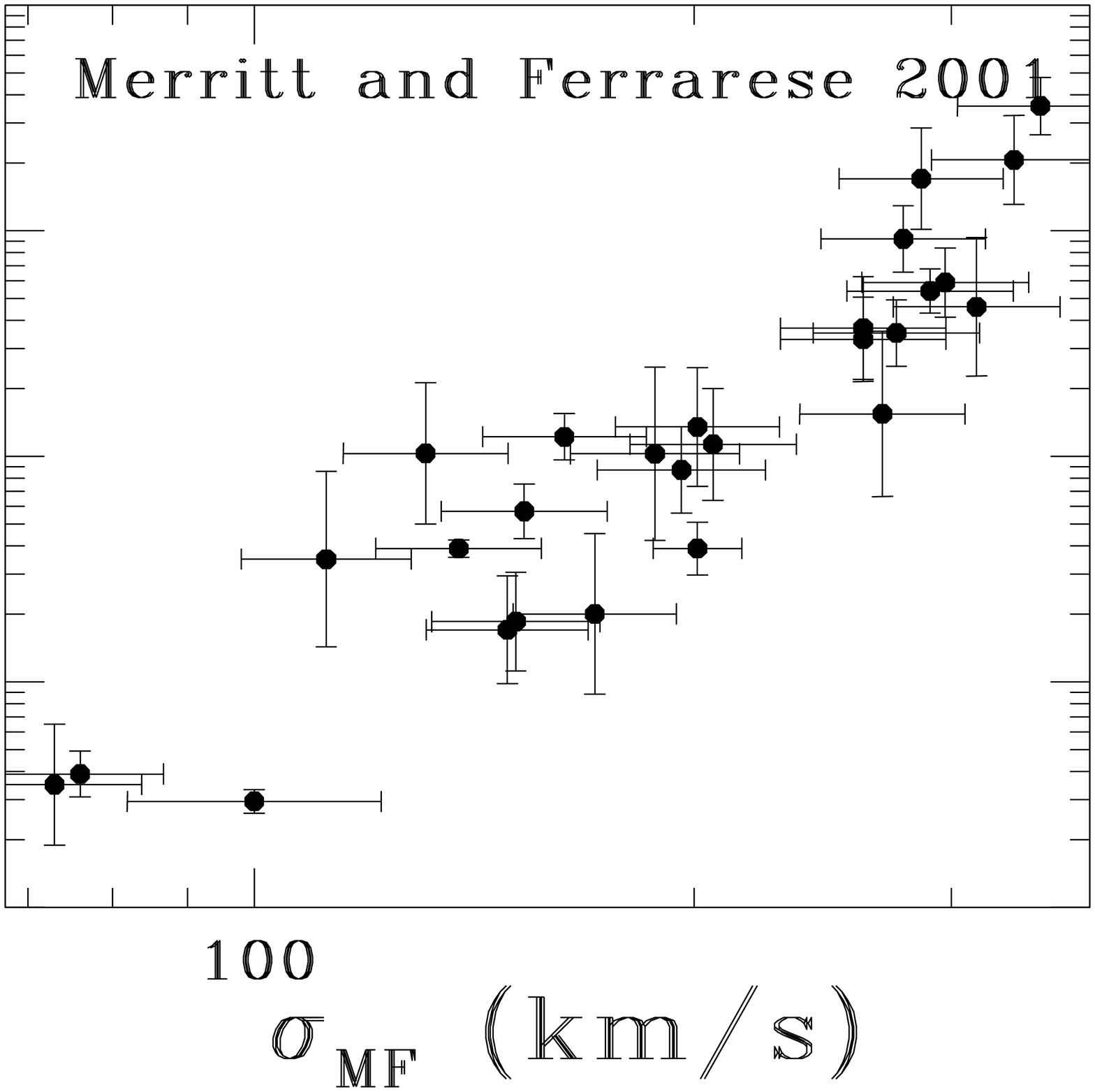}  
  \\
  \includegraphics[height=.25\textwidth]{f1a}  
  \includegraphics[height=.25\textwidth]{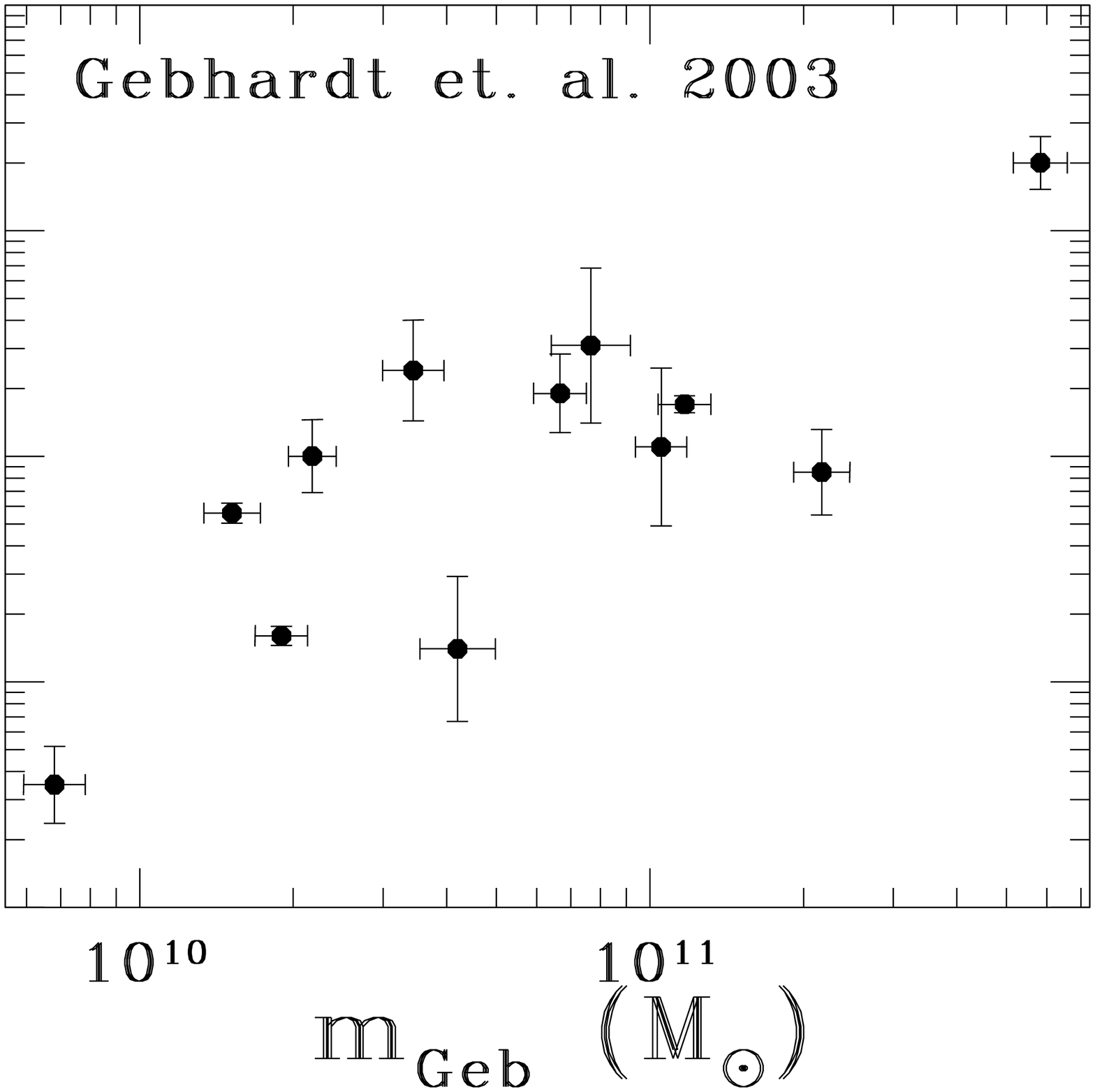}  
  \includegraphics[height=.25\textwidth]{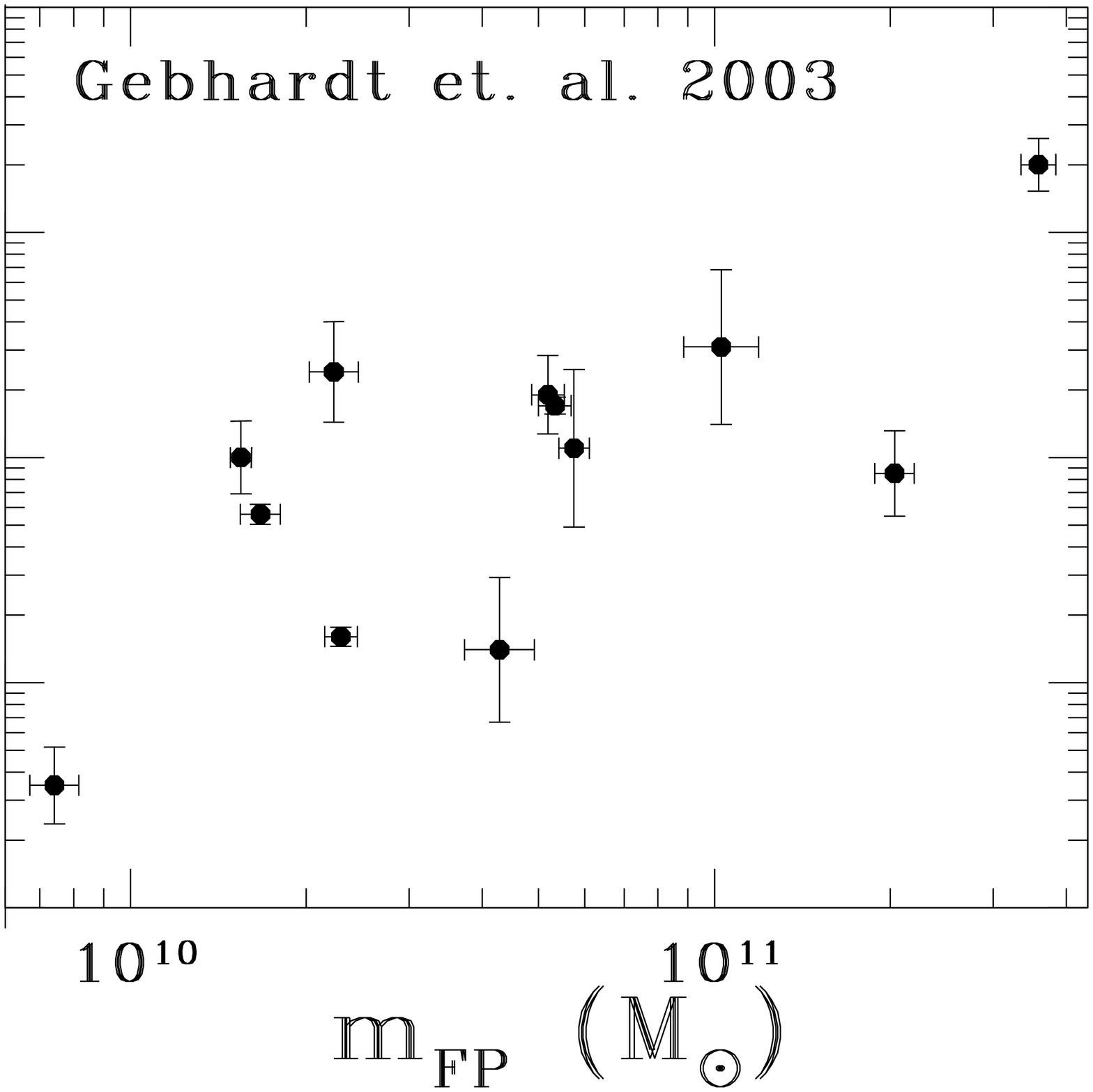}  
  \includegraphics[height=.25\textwidth]{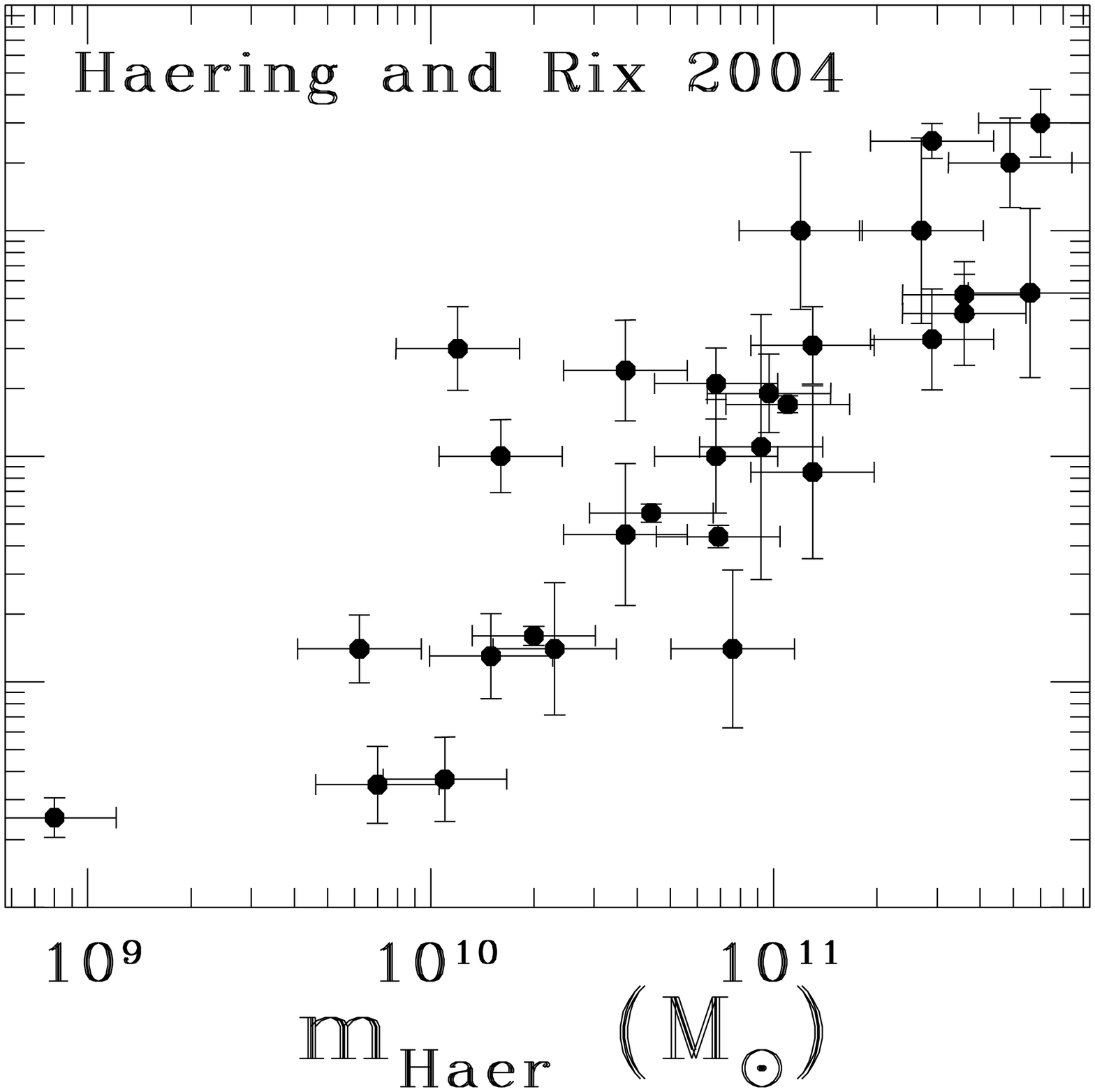}  
  \\
  \includegraphics[height=.25\textwidth]{f1a}  
  \includegraphics[height=.25\textwidth]{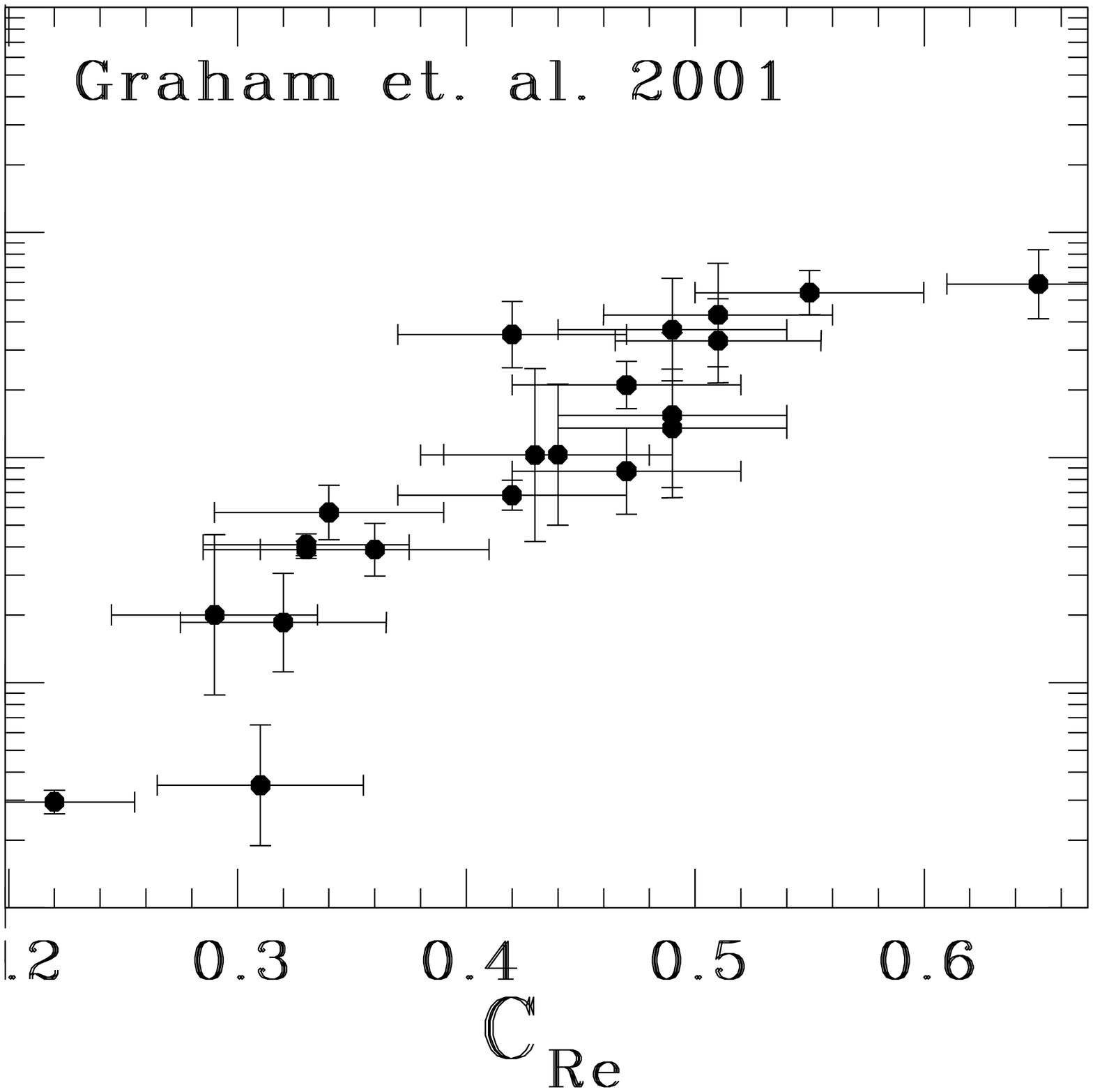}    
  \includegraphics[height=.25\textwidth]{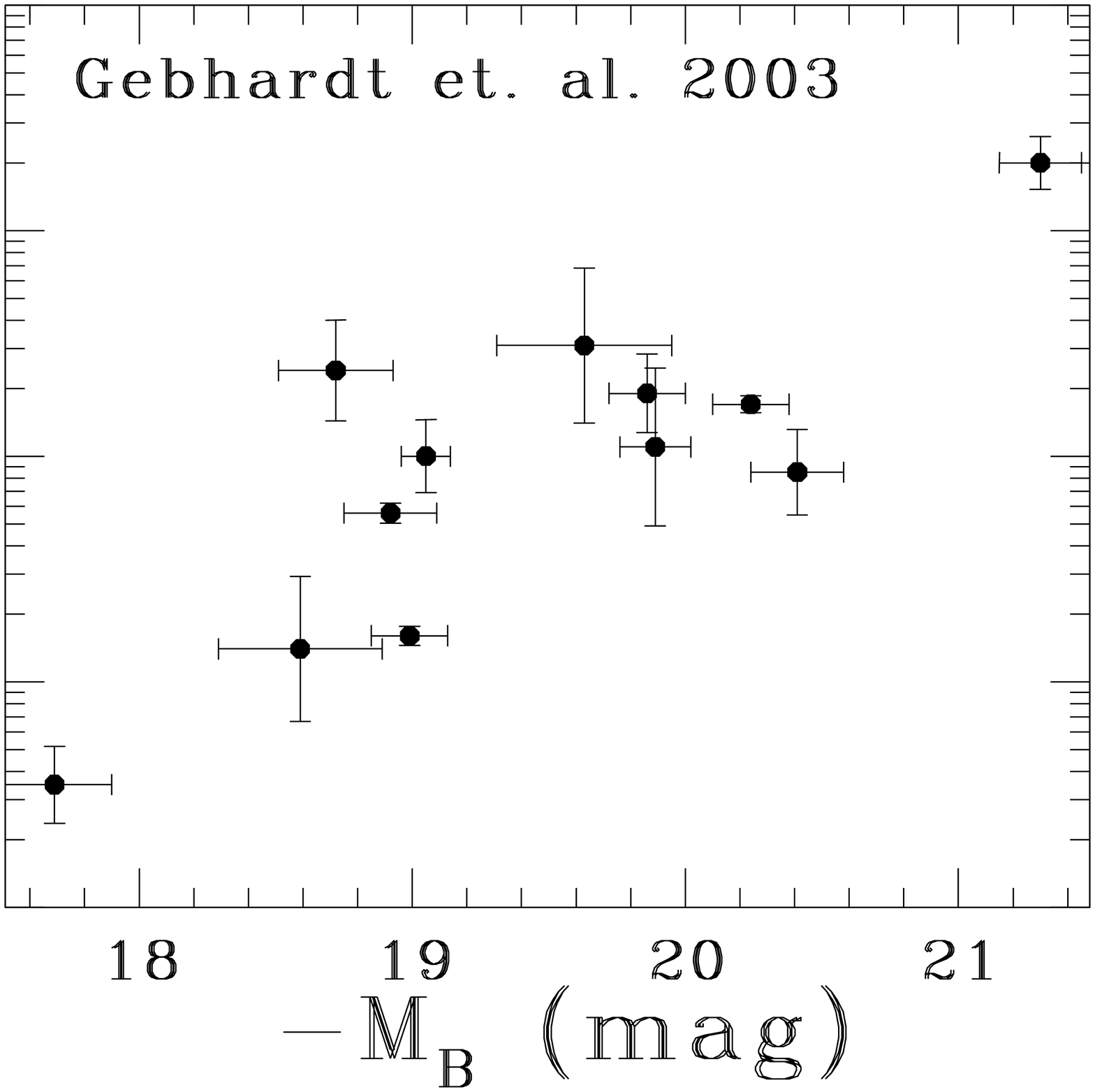}    
  \includegraphics[height=.25\textwidth]{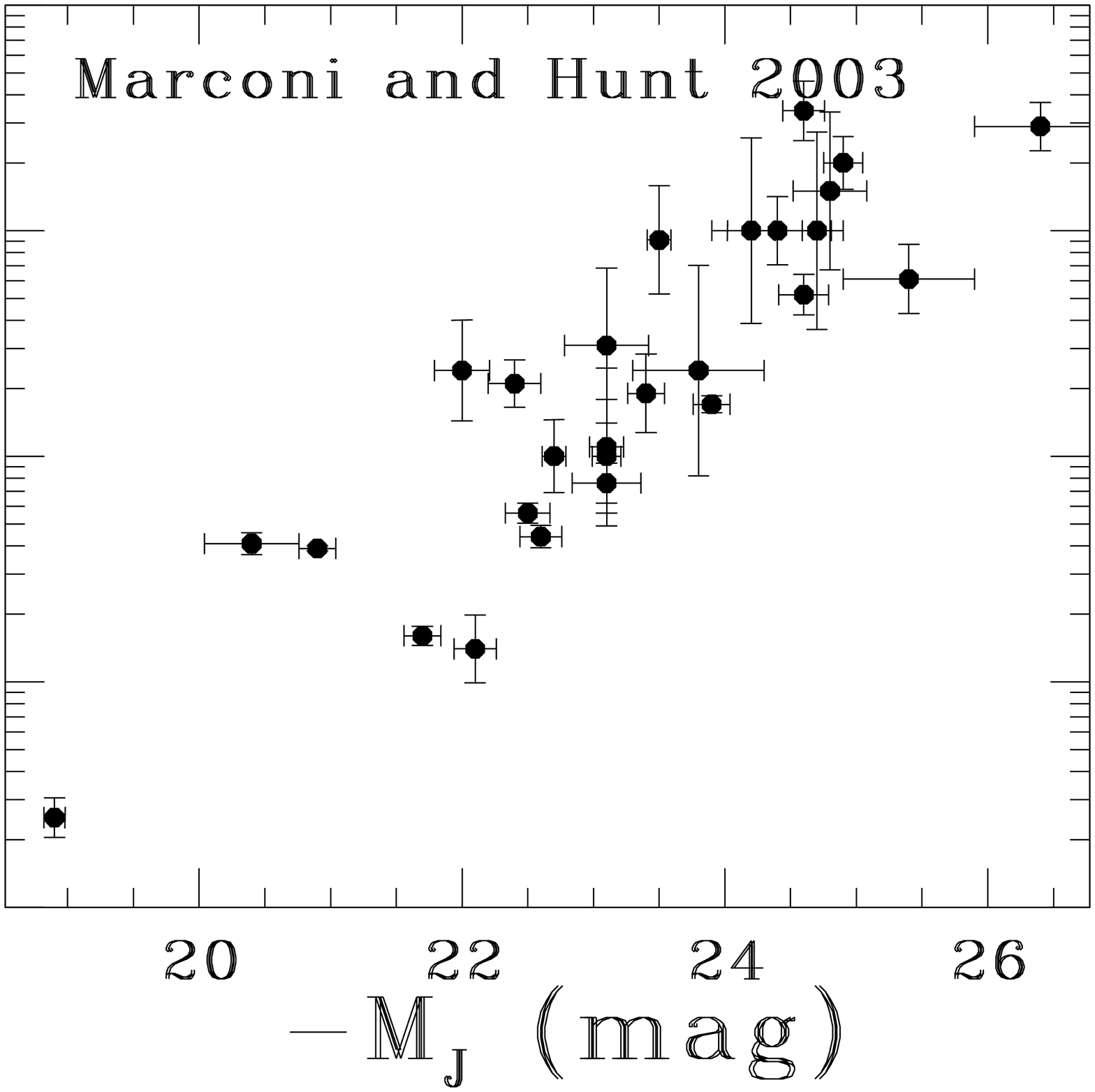}    
  \\
  \includegraphics[height=.25\textwidth]{f1a}  
  \includegraphics[height=.25\textwidth]{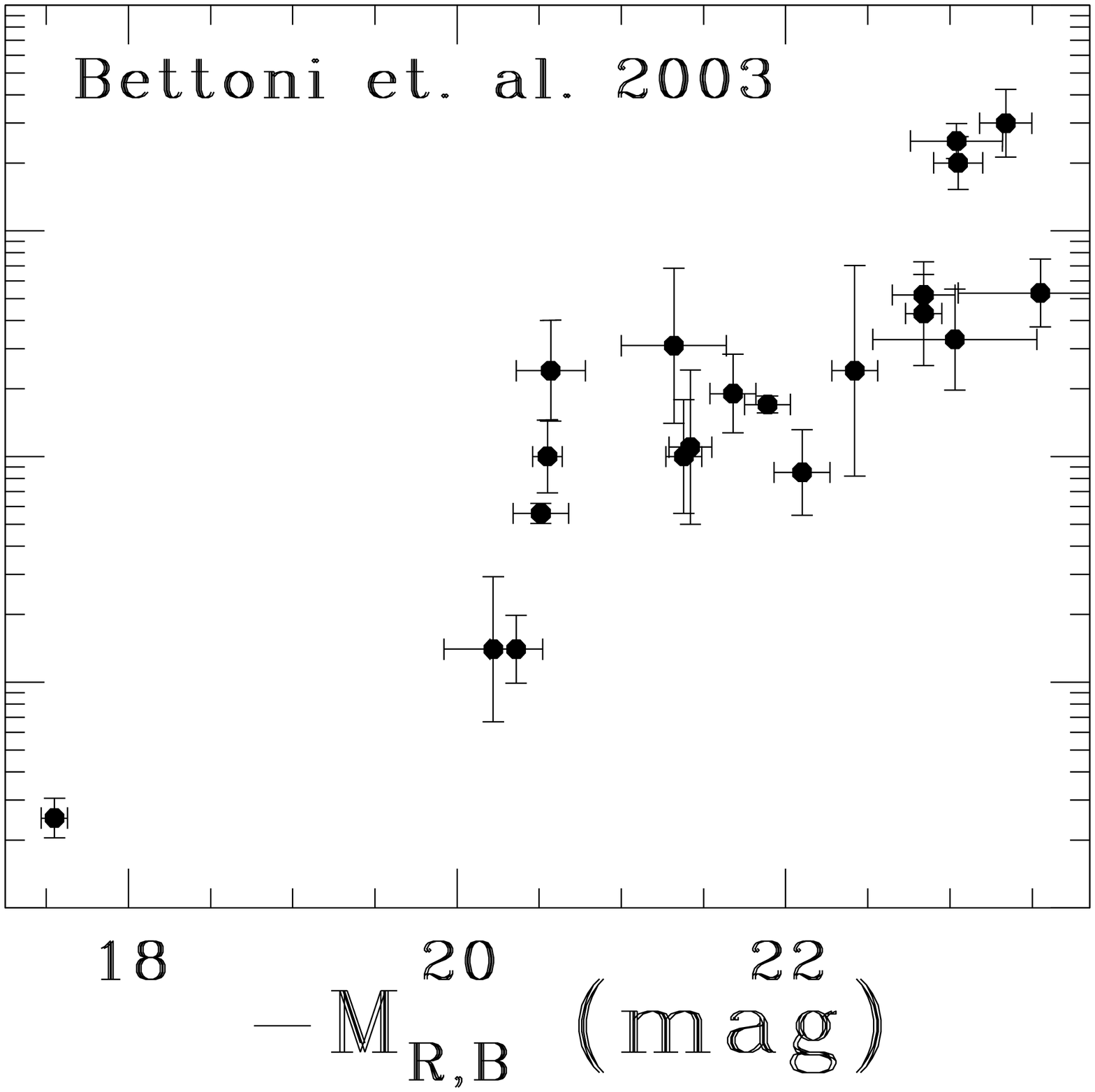}    
  \includegraphics[height=.25\textwidth]{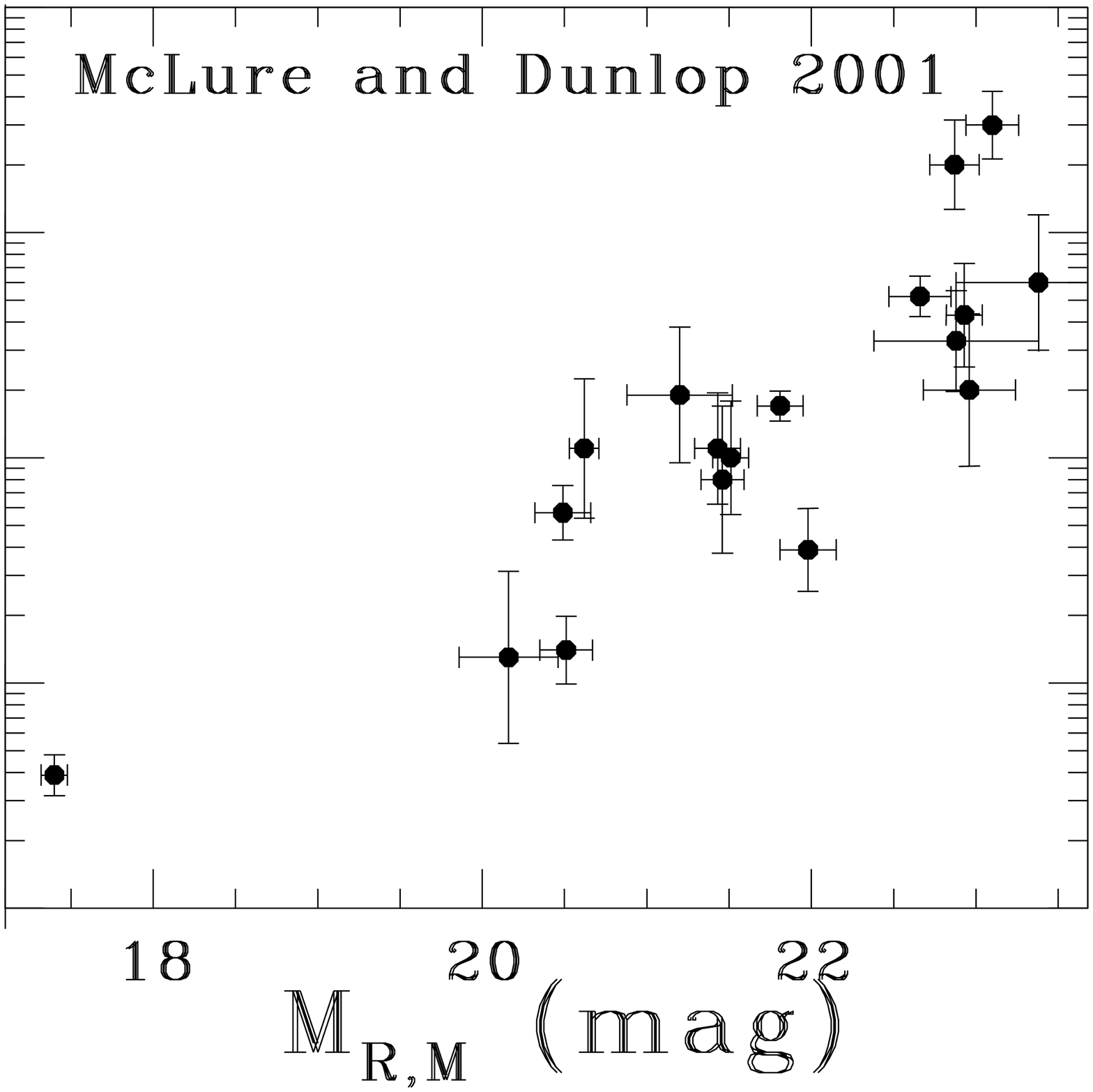}      
  \caption{BH mass plotted against different predictor variables.  See
    \S\ref{sec:data} for details.}
  \label{fig:all-data}
\end{figure}

Absolute $B$-band luminosities for the \citet{gebhardt:03}
galaxies, denoted $M_B$, are taken from the same paper.  The dominant
source of error is the random error on the distance from
\citet{tonry:01}.

The bulge masses for the \citet{gebhardt:03} galaxies are denoted
$m_{\rm bulge}$.  Absolute $B$-band luminosities and $V$-band \ml\ values
are from the same paper, distances are from \citet{tonry:01}, and
$B-V$ colors are from \citet{faber:97} or \citet{devaucouleurs:95} as indicated in
Table \ref{tab:mydata}.  The \ml\ value is derived from the same
dynamical models used to estimate BH mass.  The uncertainty in the
bulge mass comes from the published uncertainty in \ml\ combined with
the random errors in the distance from \citet{tonry:01} propagated
through the absolute magnitude using standard error propagation.

A simpler estimate of the bulge mass comes from the homology
assumption in which $m_{\rm FP} \equiv k R_e \sigma^2/G$, where $R_e$ is
the half-light radius, $\sigma$ is the velocity dispersion, $G$ is the
gravitational constant, and $k$ is a structure constant (assumed to be
the same for all galaxies in the sample) that depends on the exact
form of the mass and velocity distributions.  To compute $m_{\rm FP}$,
velocity dispersions are taken from \citet{gebhardt:03}, distances 
from \citet{tonry:01}, and $R_e$ values from \citet{faber:97},
\citet{devaucouleurs:95}, or \citet{faber:89} as indicated in Table
\ref{tab:mydata}.  The dominant error is uncertainty in the distance
propagated through $R_e$ using standard error propagation.

This core sample is enlarged by considering other data sets.  Values
and error bars are in general adopted from the published values.
\citet{tremaine:02} include the galaxies from \citet{gebhardt:03} plus
19 additional galaxies over a wide range of Hubble types.  Velocity
dispersions for these galaxies are denoted $\sigma_{\rm T}$.

\citet{merritt:01} collected BH mass estimates for 27 galaxies over a
wide range of Hubble types.  They took velocity dispersions from
the literature and corrected them to dispersions within $R_e/8$
(denoted $\sigma_{\rm MF}$) using an empirical formula.  

\citet{graham:01} collected BH mass estimates for 21 galaxies and used
galaxy light concentration, denoted $C_{\rm Re}$, as the BH mass predictor
variable.  They define $C_{\rm Re}$ to be twice the fraction of the
galaxy's light enclosed within 1/3 of the half-light radius.  They
computed $C_{\rm Re}$ by fitting a S\'ersic profile to galaxy images and
then using an analytic formula to convert the S\'ersic index $n$ to
$C_{\rm Re}$.  They estimate the error in $C_{\rm Re}$ based  on previous
experience with the fitting procedure. 
 
\citet{marconi:03} collected 37 galaxies spanning a wide range in
Hubble types and advocated the use of near-infrared luminosity to
predict BH mass.  They retrieved $J$-band apparent magnitudes from the
2MASS survey and used distances from \citet{tonry:01} or Hubble
velocities corrected for Virgo-centric infall to convert them to
absolute magnitudes.  They do not quote uncertainties in $M_J$; we
estimate them from the distance errors quoted by \citet{tonry:01}, or,
where not available, we assume 0.5 mag of uncertainty in the
distance modulus.  They exclude several galaxies, including those for
which the BH sphere of influence is less than one resolution element.
 
\citet{haering:04} used spherical, isotropic Jeans modeling to
estimate the bulge masses of 10 galaxies and drew bulge mass
estimates from the literature for an additional 20 galaxies,
including 12 from \citet{magorrian:98}.  They used BH mass
estimates mostly from \citet{tremaine:02} to argue that the
BH-mass-to-bulge-mass correlation has similar residual variance to the
other correlations under consideration here.  Their bulge
mass estimates are denoted $m_{\rm Haer}$.

\citet{mclure:02} drew the E-type galaxies from the sample published
in \citet{kormendy:01} (discarding two because the uncertainties in
the BH masses were larger than for the rest of the sample) and used
$B$-band luminosities from \citet{faber:97}, $V$-band luminosities
from \citet{merritt:01}, and average color corrections from
\citet{fukugita:95} to get $R$-band luminosities for 18 galaxies.  The
dominant error comes from the uncertainty in the distance from
\citet{tonry:01}.  Where distances based on surface brightness
fluctuations are not available, we assume 0.5 mag uncertainty in the
distance modulus.  They argued that the problem with the
BH-mass-to-bulge-luminosity relation was simply bulge-disk
decomposition, and they claimed that removing spiral and S0 galaxies resulted in
a very good correlation.  They do not provide a data table, but they
do describe how they selected their sample.  We attempt to reconstruct
it.  Their $R$-band absolute magnitudes are denoted $M_{R, \rm M}$, where
the M stands for McLure.

This notation is necessary because \citet{bettoni:03} also used
$R$-band luminosities in a very similar sample.  They also drew only
E-type galaxies from \citet{kormendy:01}, resulting in a sample of 20
galaxies.  They drew $B$-band luminosities from \citet{faber:97} for
all but three galaxies, for which they obtained $B$-band luminosities
from \citet{devaucouleurs:95}.  To get $R$-band luminosities they used colors from
\citet{prugniel:00} or \citet{devaucouleurs:95}.   Their $R$-band
magnitudes are denoted $M_{R, \rm B}$.

To simplify the analysis we make all error bars symmetric about the
preferred value by averaging the size of the upper and lower
1-$\sigma$ error bars so that $x^{+h}_{-l}$ becomes $x \pm
(h+l)/2$.   This simplification should not greatly affect
the result, since most of the confidence regions are reasonably
symmetric in log space.

\section{Results}
\label{sec:results}

\subsection{Theorist's Question}
\label{sec:results-theorist}
Figure \ref{fig:all-data} shows the raw data for the 11 global galaxy
quantities described in \S\ref{sec:data}.  We apply the method
described in \S\ref{sec:method-theorist} to all of these
variables in order to infer the residual variance associated with each
linear fit.

Above, we noted that one's choice about how to include residual
variance---whether in $x$ or in $y$---affects estimates of the slope
of a given correlation.  For the galaxies listed in
\citet{tremaine:02}, the slope is $4.59 \pm 0.34$ or $4.10 \pm 0.30$
depending on whether the residual variance is
assigned to $x$ or $y$, respectively.  For \citet{merritt:01} galaxies, the
slope is $4.54 \pm 0.38$ or $4.42 \pm 0.37$.  One might choose any one
of these numbers as a point of comparison for a theoretical
model, depending on the content of the model. 

\begin{figure}[htbp]
  \includegraphics[height=.5\textheight]{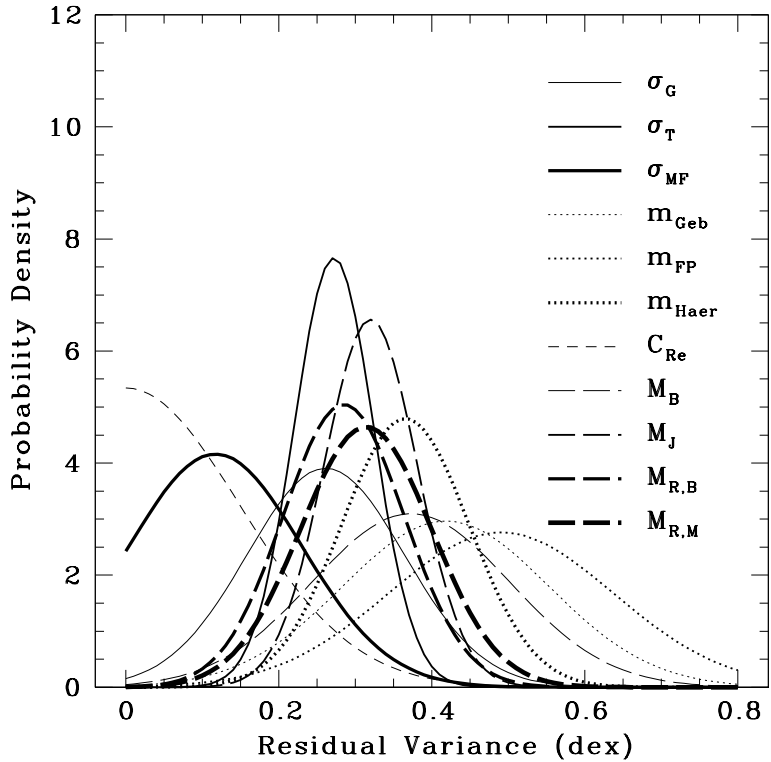}
  \caption{Probability distributions for residual variance for 11
    BH mass predictor variables.  Residual variance is the variance
    that is left over after observational errors are accounted for; a
    smaller residual variance means the correlation is intrinsically
    tighter.  For most of the correlations, the available data do not
    constrain the residual variance well enough to make strong
    statements about which correlation is preferred over the others.
    The only 3 $\sigma$ statement that can be made is that $C_{\rm Re}$ is
    better than $m_{\rm FP}$, $m_{\rm Haer}$, and $M_J$, and this depends
    critically on the assumption that errors on all predictor
    variables are accurately estimated.  The galaxy light
    concentration ($C_{\rm Re}$) and velocity dispersion as measured by
    \citet{merritt:01} have residual variance distributions that are
    peaked at lower values than the others.  Both relations are
    consistent with zero residual variance.  Figure 3 shows that they
    are also the only two relations with predictive uncertainties in
    $M_{BH}$ dominated by measurement error rather than residual
    variance.  If these relations truly are better than the others,
    then more accurate measurements of the predictor variable will
    allow very good BH mass inferences.}
  \label{fig:all-dists}
\end{figure}

The most straightforward comparison among the different correlations
is obtained under the assumption that the residual variance is always
in the BH mass, so that it has the same units in all cases.  Figure
\ref{fig:all-dists} shows the probability distribution of possible
\cv\ values given the data sets published for each correlation under
consideration.  We estimate the 1 $\sigma$ error on \cv\ as described
in \S\ref{sec:method-theorist} and then plot a Gaussian distribution
of the absolute value of \cv\ in order to emphasize the extent to
which the allowed ranges overlap.  These distributions of \cv\ are most
relevant for answering the Theorist's Question, which asks what
quantity is most closely associated with BH mass for the purpose of
constructing theoretical models.  The best intrinsically correlated
quantity will have the distribution of \cv\ closest to zero.

The differences between the various correlations are not highly
significant, as most distributions overlap substantially.  
This indicates that the available data do not
constrain the residual variance well enough to make strong statements
about which correlation is intrinsically tighter than the others.  
The galaxy light concentration $C_{\rm Re}$ and the velocity dispersion
$\sigma_{\rm MF}$ as measured by \citet{merritt:01} show residual variance
distributions that are peaked at lower values than the others, and
each is consistent with zero. At face value, this means that these two 
correlations are better than the others, provided that their error
estimates are accurate.  

There seems to be a slight systematic trend among the types of
quantities---galaxy light concentration has the smallest residual
variance, followed by the three  velocity dispersions, then the four
luminosities, and finally the three masses.  However, it is difficult
to make much of this, since the individual distributions overlap
significantly, and many of the different samples are correlated because
they have many galaxies in common. 

If one omits the galaxies for which the BH sphere of influence is unresolved
according to \citet{marconi:03}, all of the curves become less sharply
peaked, because the samples have gotten smaller.  This effect is
largest for $C_{\rm Re}$, where the 1 $\sigma$ upper limit goes from 0.15
to 0.23.  This is a somewhat larger change than one would expect just
from the change in sample size, indicating that the unresolved
galaxies happen to lie very close to the fitted line for this
relation.

The question of which correlation is intrinsically tightest will remain 
open until more accurate data become available or the number of
objects increases dramatically.  There is some indication that
reverberation mapping may become useful in constraining the residual
variance of the BH mass correlations by greatly increasing the number
of objects with BH mass estimates of sufficient precision
\citep{onken:04}. 

If one is willing to view the many experiments undertaken in the
literature involving various subsamples, various galaxy variables, and
various corrections as a way of learning about the distribution of
\cv\ via an empirical bootstrap-like error estimate, the conclusion
agrees with our result nicely: quoted numbers for \cv\ 
range from $\sim$0.2 to $\sim$0.5 dex under a wide variety of
assumptions.

In spite of the fact that the Theorist's Question is not answerable at
this time, we include a compilation of our fits to all of the data
sets assuming residual variance in both the $x$- and $y$-coordinates  
in Tables \ref{tab:fits-one} and \ref{tab:fits-two}, for
the purpose of comparing the slopes of the many theoretical models to
the observations. 

Table \ref{tab:correlations} provides Pearson and Spearman correlation
coefficients along with confidence regions.  The Pearson and Spearman
values do not differ markedly except for the smallest data sets, which
contain 12 galaxies.  For these data sets, the confidence regions
reveal that the coefficients are essentially unconstrained.

\subsection{Observer's Question}
\label{sec:results-observer}
Next we consider which correlation best predicts the BH masses of
galaxies for which there is no direct measurement.  Such correlations
are in great demand to answer a variety of astrophysical questions
\citep[e.g.,][]{salucci:99,aller:02,yu:02,barth:03,marconi:04}.  Here
the relevant quantity is the predictive power of each correlation,
which we explore with our poor man's predictive distribution
defined in \S\ref{sec:method-observer}.  Inherent in this approach is
the assumption that future data sets used to predict BH masses will
have the same predictor measurement errors as those discussed here.
The method can handle other assumptions, but we do not pursue them
here.

\begin{figure}[htbp]
  \includegraphics[height=.5\textheight]{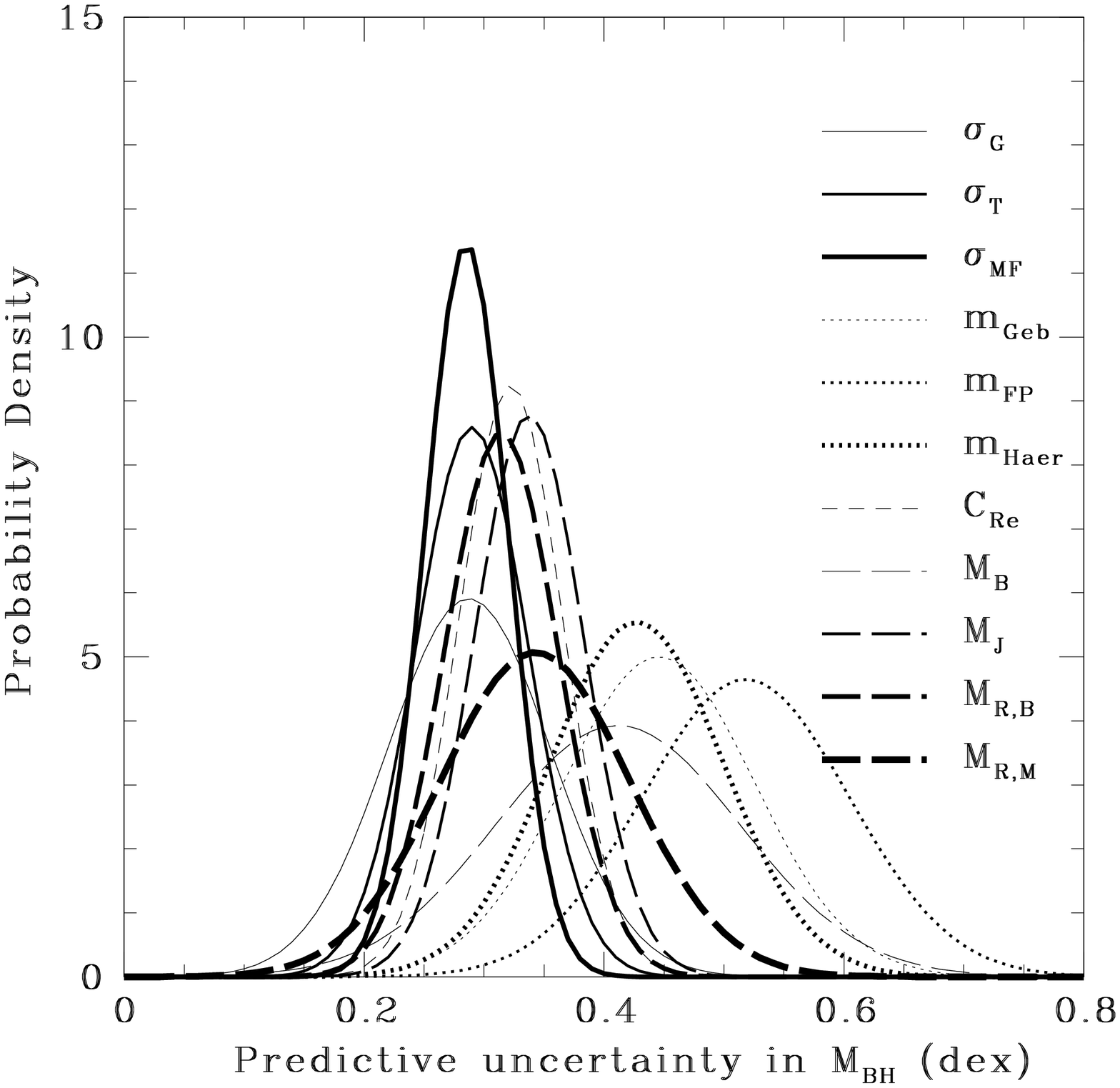}
  \caption{Expected uncertainty in inferred BH mass for a new
    measurement of the predictor variable which results in a BH mass
    near $10^8$ $M_\odot$ (near the center of the observed range).
    Near the edges of the range, the
    expected uncertainty goes up somewhat because the slope of each
    relation in not known precisely.  The peak of each distribution
    comes from equation \ref{eq:predictive}, and the width is
    determined by 100 bootstrap Monte Carlo samples on each data set
    as described in \S\ref{sec:results-observer}.
    If a curve is peaked in the same place as in Figure
    \ref{fig:all-dists}, then the predictive uncertainty is dominated
    by residual variance; if a curve moves significantly, then the
    predictive uncertainty is dominated by measurement error.
    Galaxy light concentration $C_{\rm Re}$ and velocity dispersion
    $\sigma_{\rm MF}$ are dominated by measurement error, while the others
    are dominated by residual variance.  The width of each curve
    (the error on the error) is dominated by uncertainty in the
    residual variance in all cases.  The only 3 $\sigma$ statement
    that can be made is that $\sigma_{\rm T}$, $\sigma_{\rm MF}$, and
    $M_{R, \rm B}$ are better than $m_{\rm FP}$.}
  \label{fig:predictive}
\end{figure}

For each predictor variable, we consider a hypothetical new
measurement with observational error equal to the mean of the absolute
value of the observational errors for existing measurements of that
predictor variable.  All of the residual variance contributes to
uncertainty in the BH mass, so we do the fits assuming that the
residual variance is in the $y$-coordinate.  Hypothetical $x$-values
are chosen to lie near the center of the range of observed values.
For velocity dispersions, $x=190 \kms$, for bulge masses
$x=6\times10^{10}$ \msun, for galaxy light concentration $x=0.435$, for
$B$-band magnitude $x=-20.4$, for $J$-band magnitude $x=-23.4$, and
for $R$-band magnitude $x=-22.8$.

Given the hypothetical new measurement, we seek our best estimate
of the error in the predicted BH mass and the uncertainty in this
quantity.  The error itself is obtained from our poor man's
predictive probability distribution.  To obtain the ``error on the
error'' we produce 100 bootstrap Monte Carlo data sets from each data
set by selecting data points at random with replacement.  For each
synthetic data set we fit a line and then evaluate the predictive
uncertainty in the BH mass.  The variance of the predictive
uncertainties associated with each of the 100 synthetic data sets
gives us the error on the error.

In Figure \ref{fig:predictive}, the peak of each curve is the most
likely value of the uncertainty in a new BH mass measurement.  The
width of each curve shows the uncertainty in this quantity determined
as described above by Bootstrap Monte Carlo simulations.  It turns out
that this is dominated by uncertainty in \cv\ resulting from small
data sets for all predictors.

The conclusion from Figure \ref{fig:predictive} is that nearly all of
the curves overlap significantly, and this turns out to be  because
the residual variance is not well constrained.  Most of the curves
are similar between Figures \ref{fig:all-dists} and
\ref{fig:predictive}, indicating that the predictive uncertainty of
most of the relations is dominated by residual variance. 

Two correlations, with $C_{\rm Re}$ and $\sigma_{\rm MF}$, move significantly
between the two figures, meaning that the predictive uncertainty is
dominated by observational error.  As noted in
\S\ref{sec:method-observer}, this is consistent with the possibility
that the estimated error bars for these two quantities are too large.
If the error bars {\em are} accurate, then more accurate measurements
of either of these two quantities will offer excellent BH mass
inferences.  Additional data will be able to test whether or not this
is the case.

Ideally, a comparison of different predictors of BH mass would be
done using the same galaxy sample for each of the different
predictors.  However, the uncertainty associated with small sample
sizes dominates both the uncertainty in the residual variance and
the uncertainty in the predictive error in BH mass for the
various relations.  The sizes of the data sets must be larger before
it becomes profitable to remove galaxies that do not appear in all of
the data sets.

\section{Conclusions}
\label{sec:conclusions}
We have constructed a framework with which to compare the correlations
between BH mass and various global galactic properties.  For each
galaxy quantity, we estimate the preferred values and uncertainties in
the slope, offset, and residual variance of the relation.  Estimating
the error on the residual variance is particularly important, since
handling it systematically makes clear that it is not
well enough constrained to draw strong distinctions between the power
of various proposed correlations.  We assume that all of the relations
are power laws.  It may be possible to reduce the residual variance of
a relation at the expense of additional model parameters.

Of the BH mass predictor variables considered here, we find that 
$\sigma_{\rm MF}$ and $C_{\rm Re}$ seem to exhibit the smallest residual
variance, and thus the tightest intrinsic correlations with the BH
mass, but the only 3 $\sigma$ statement that can be made is that
$C_{\rm Re}$ is better than $M_J$, $m_{\rm Haer}$, and $m_{\rm FP}$.  Any
such claims about the residual variance also depend critically on the
assumed measurement errors.  

Concerning the use of these correlations to infer BH mass, we find
that again the existing data sets do not offer a clear favorite.  Some
correlations may be marginally better than others, but the only
3 $\sigma$ statement is that $\sigma_{\rm T}$, $\sigma_{\rm MF}$, and
$M_{R, \rm B}$ are preferable to $M_{\rm FP}$.  There does not seem to be
a profound difference arising from the different ways of measuring
velocity dispersion.  

The only two quantities for which BH mass inferences are dominated by
measurement error are $C_{\rm Re}$ and $\sigma_{\rm MF}$.  If the error bars
on these quantities are correct, more accurate data should allow
excellent BH mass inferences.  

The murkiness of the present situation is due almost entirely to the
limited number of galaxies for which reliable BH mass measurements are
available.  Consistently handling the uncertainty in the residual
variance due to the finite sample size reveals that there are simply
too few data to make strong statements.

We caution that all of these conclusions depend critically on the
accuracy of the published error estimates in all quantities under
consideration.  The common practice of adopting conservatively large
error bars with the intention of ensuring that the true value lies
within the quoted range is undesirable in this case, since it
artificially suppresses the amount of residual variance required by
the fit.

\acknowledgments

G. S. N. was supported by the Krell Institute through the
Computational Science Graduate Fellowship Program.  This research has
been partly supported by ISF 213/02 and NASA ATP NAG5-8218.  S. M. F. would
like to acknowledge the support of a Visiting Miller Professorship at
UC Berkeley.  The authors would like to thank Scott Tremaine and
the anonymous referee for comments that substantially improved the
paper.

\bibliographystyle{apj}
\bibliography{ms}

\begin{deluxetable}{rrrrrrrr}
\tablenum{1}
\tablecolumns{8} 
\tablewidth{0pc} 
\tablecaption{BH mass predictor variables $m_{\rm FP}$ and $m_{\rm bulge}$
  for galaxies from \citet{gebhardt:03}}
\tablehead{
  \colhead{Name} & \colhead{Distance (Mpc)} & \colhead{$R_e$ (kpc)} & \colhead{$R_e$ ref.} & \colhead{$m_{\rm FP}$ (\msun)} &
  \colhead{$B-V$ (mag)} & \colhead{$B-V$ ref.} & \colhead{$m_{\rm bulge}$ (\msun)}}
\startdata 
N0821 & $ 24.1 \pm 4.1 $ &11.63 & 2 & $ 2.03^{+0.32}_{-0.28} \times 10^{11} $ & 0.93 & 4 & $ 2.17^{+0.47}_{-0.39} \times 10^{11} $\\ 
N2778 & $ 22.9 \pm 6.9 $ & 3.50 & 2 & $ 4.29^{+1.28}_{-0.98} \times 10^{10} $ & 0.91 & 4 & $ 4.20^{+1.56}_{-1.14} \times 10^{10} $\\ 
N3377 & $ 11.2 \pm 1.0 $ & 1.84 & 1 & $ 1.55^{+0.13}_{-0.12} \times 10^{10} $ & 0.84 & 5 & $ 2.18^{+0.31}_{-0.27} \times 10^{10} $\\ 
N3384 & $ 11.6 \pm 1.6 $ & 2.80 & 2 & $ 2.29^{+0.30}_{-0.26} \times 10^{10} $ & 0.91 & 5 & $ 1.90^{+0.36}_{-0.30} \times 10^{10} $\\ 
N3608 & $ 22.9 \pm 3.2 $ & 3.91 & 1 & $ 5.19^{+0.67}_{-0.59} \times 10^{10} $ & 0.98 & 5 & $ 6.67^{+1.25}_{-1.05} \times 10^{10} $\\ 
N4291 & $ 26.2 \pm 8.4 $ & 4.38 & 2 & $ 1.03^{+0.33}_{-0.25} \times 10^{11} $ & 0.93 & 4 & $ 7.66^{+3.05}_{-2.18} \times 10^{10} $\\ 
N4473 & $ 15.7 \pm 2.0 $ & 3.98 & 2 & $ 5.75^{+0.69}_{-0.61} \times 10^{10} $ & 0.92 & 4 & $ 1.05^{+0.19}_{-0.16} \times 10^{11} $\\ 
N4564 & $ 15.0 \pm 2.5 $ & 1.59 & 1 & $ 1.67^{+0.27}_{-0.23} \times 10^{10} $ & 0.98 & 5 & $ 1.52^{+0.33}_{-0.27} \times 10^{10} $\\ 
N4649 & $ 16.8 \pm 2.5 $ & 6.04 & 1 & $ 3.58^{+0.50}_{-0.44} \times 10^{11} $ & 0.99 & 5 & $ 5.83^{+1.15}_{-0.96} \times 10^{11} $\\ 
N4697 & $ 11.7 \pm 1.6 $ & 4.25 & 1 & $ 5.33^{+0.69}_{-0.61} \times 10^{10} $ & 0.95 & 5 & $ 1.17^{+0.22}_{-0.18} \times 10^{11} $\\ 
N5845 & $ 25.9 \pm 5.4 $ & 1.02 & 3 & $ 2.23^{+0.45}_{-0.37} \times 10^{10} $ & 0.97 & 5 & $ 3.44^{+0.90}_{-0.71} \times 10^{10} $\\ 
N7457 & $ 13.2 \pm 2.8 $ & 4.13 & 2 & $ 7.41^{+1.48}_{-1.24} \times 10^{9} $  & 0.83 & 5 & $ 6.81^{+1.78}_{-1.41} \times 10^{9} $ \\ 
\enddata 
\tablecomments{All distances are from  \citet{tonry:01}.
  Half-light radius ($R_e$) sources are: (1) \citealt{faber:97}; (2)
  \citealt{devaucouleurs:95}; (3) \citealt{faber:89}.  $B-V$ color sources are: (4)
  \citealt{faber:97}; (5) \citealt{devaucouleurs:95}.}
\label{tab:mydata} 
\end{deluxetable} 

\begin{deluxetable}{lrrrrrrr}
  \tablecolumns{6} 
  \tablewidth{0pt}
  \tablenum{2}
  \tablecaption{Fit Parameters for Residual Variance in the $x$-Coordinate}
  \tablehead{\colhead{Quantity} & \colhead{$\alpha$} &
    \colhead{$\beta$} & \colhead{$\cv$} & \colhead{Zero point} & \colhead{$N$}}
  \startdata
  $\sigma_{\rm G}$    & 8.168 $\pm$ 0.099 & 4.18  $\pm$ 0.57  & 0.066  $\pm$ 0.023  & 200 \kms        & 12 \\
  $\sigma_{\rm T}$    & 8.155 $\pm$ 0.062 & 4.59  $\pm$ 0.34  & 0.063  $\pm$ 0.012  & 200 \kms        & 31 \\
  $\sigma_{\rm MF}$ & 8.117 $\pm$ 0.066 & 4.54  $\pm$ 0.38  & 0.030  $\pm$ 0.020  & 200 \kms        & 27 \\
  $m_{\rm Geb}$     & 8.38  $\pm$ 0.16  & 1.55  $\pm$ 0.33  & 0.329  $\pm$ 0.088  & $10^{11}$ \msun & 12 \\
  $m_{\rm FP}$      & 8.65  $\pm$ 0.20  & 1.92  $\pm$ 0.42  & 0.343  $\pm$ 0.086  & $10^{11}$ \msun & 12 \\
  $m_{\rm Haer}$    & 8.348 $\pm$ 0.096 & 1.38  $\pm$ 0.15  & 0.293  $\pm$ 0.065  & $10^{11}$ \msun & 30 \\
  $C_{\rm Re}$      & 8.438 $\pm$ 0.078 & 6.56  $\pm$ 0.57  & $\le$ 0.023             & 0.5             & 21 \\   
  $-M_B$        & 8.39  $\pm$ 0.14  & 0.83  $\pm$ 0.15  & 0.52   $\pm$ 0.16   & -20             & 12 \\
  $-M_J$        & 8.190 $\pm$ 0.081 & 0.524 $\pm$ 0.056 & 0.68   $\pm$ 0.12   & -23             & 27 \\
  $-M_{R, \rm B}$    & 8.456 $\pm$ 0.084 & 0.554 $\pm$ 0.059 & 0.55   $\pm$ 0.14   & -22             & 20 \\
  $-M_{R, \rm M}$    & 8.37  $\pm$ 0.10  & 0.537 $\pm$ 0.072 & 0.65   $\pm$ 0.16   & -22             & 18 \\
  \enddata
  \label{tab:fits-one} 
\end{deluxetable}

\begin{deluxetable}{lrrrrrrr}
  \tablecolumns{6} 
  \tablewidth{0pt}
  \tablenum{3}
  \tablecaption{Fit Parameters for Residual Variance in the $y$-Coordinate}
  \tablehead{\colhead{Quantity} & \colhead{$\alpha$} &
    \colhead{$\beta$} & \colhead{$\cv$} & \colhead{Zero point} & \colhead{N}}
  \startdata
  $\sigma_{\rm G}$    & 8.143 $\pm$ 0.094 & 3.69  $\pm$ 0.51  & 0.261 $\pm$ 0.091 & 200 \kms        & 12 \\
  $\sigma_{\rm T}$    & 8.133 $\pm$ 0.059 & 4.10  $\pm$ 0.30  & 0.272 $\pm$ 0.050 & 200 \kms        & 31 \\
  $\sigma_{\rm MF}$ & 8.116 $\pm$ 0.065 & 4.42  $\pm$ 0.37  & 0.133 $\pm$ 0.090 & 200 \kms        & 27 \\
  $m_{\rm Geb}$     & 8.24  $\pm$ 0.13  & 1.05  $\pm$ 0.22  & 0.42  $\pm$ 0.12  & $10^{11}$ \msun & 12 \\
  $m_{\rm FP}$      & 8.34  $\pm$ 0.15  & 1.07  $\pm$ 0.25  & 0.49  $\pm$ 0.13  & $10^{11}$ \msun & 12 \\
  $m_{\rm Haer}$    & 8.291 $\pm$ 0.086 & 1.13  $\pm$ 0.12  & 0.365 $\pm$ 0.080 & $10^{11}$ \msun & 30 \\
  $C_{\rm Re}$      & 8.438 $\pm$ 0.078 & 6.56  $\pm$ 0.57  &  $\le$ 0.15           & 0.5             & 21 \\
  $-M_B$        & 8.28  $\pm$ 0.12  & 0.63  $\pm$ 0.12  & 0.37  $\pm$ 0.12  & -20             & 12 \\
  $-M_J$        & 8.190 $\pm$ 0.074 & 0.426 $\pm$ 0.046 & 0.320 $\pm$ 0.058 & -23             & 27 \\
  $-M_{R, \rm B}$    & 8.430 $\pm$ 0.080 & 0.488 $\pm$ 0.051 & 0.285 $\pm$ 0.075 & -22             & 20 \\
  $-M_{R, \rm M}$    & 8.325 $\pm$ 0.092 & 0.441 $\pm$ 0.058 & 0.314 $\pm$ 0.081 & -22             & 18 \\
  \enddata
  \label{tab:fits-two} 
\end{deluxetable}

\begin{deluxetable}{lrrrrrrr}
  \tablecolumns{6} 
  \tablewidth{0pt}
  \tablenum{4}
  \tablecaption{Correlation Coefficients with Confidence Regions.}
    
  \tablehead{\colhead{Quantity} & \colhead{N} & \colhead{Pearson $r$} &
  \colhead{Confidence Region} & \colhead{Spearman $r$} & 
  \colhead{Confidence Region}}
  \startdata
  $\sigma_{\rm G}$    & 12 & 0.89 & [-0.62, 0.998] & 0.85 & [-0.72, 0.998] \\
  $\sigma_{\rm T}$    & 31 & 0.92 & [0.82, 0.96]  & 0.92 & [0.82, 0.96] \\
  $\sigma_{\rm MF}$ & 27 & 0.92 & [0.81, 0.97]  & 0.92 & [0.82, 0.97] \\
  $m_{\rm Geb}$     & 12 & 0.78 & [-0.81, 0.997] & 0.63 & [-0.89, 0.99] \\
  $m_{\rm FP}$      & 12 & 0.71 & [-0.85, 0.995] & 0.59 & [-0.90, 0.99] \\
  $m_{\rm Haer}$    & 30 & 0.84 & [0.67, 0.92]  & 0.84 & [0.68, 0.93] \\
  $C_{\rm Re}$      & 21 & 0.89 & [0.67, 0.97]  & 0.93 & [0.76, 0.98] \\
  $-M_B$        & 12 & 0.81 & [-0.77, 0.997] & 0.60 & [-0.90, 0.99] \\
  $-M_J$        & 27 & 0.88 & [0.73, 0.95]  & 0.86 & [0.69, 0.94] \\
  $-M_{R, \rm B}$    & 20 & 0.89 & [0.62, 0.97]  & 0.88 & [0.61, 0.97] \\
  $-M_{R, \rm M}$    & 18 & 0.86 & [0.46, 0.97]  & 0.85 & [0.41, 0.97] \\
  \enddata
  \label{tab:correlations} 
  \tablecomments{Confidence regions are 68\% regions computed assuming
    that $\tanh^{-1}r$ is distributed normally,
    as described in \citet[][p. 129]{cowan:98}}
\end{deluxetable}
\end{document}